\documentclass[authordraft]{acmart}
\AtBeginDocument{%
  }

\setcopyright{acmlicensed}
\copyrightyear{2018}
\acmYear{2018}
\acmDOI{XXXXXXX.XXXXXXX}
\acmConference[Conference acronym 'XX]{Make sure to enter the correct
  conference title from your rights confirmation email}{June 03--05,
  2018}{Woodstock, NY}
\acmISBN{978-1-4503-XXXX-X/18/06}

\begin{document}

\title{Demographic Benchmarking: Bridging Socio-Technical Gaps in Bias Detection}

\author{Gemma Galdon Clavell}
\email{gemma@eticas.ai}
\orcid{0000-0002-6481-0833}
\affiliation{%
  \institution{Eticas AI}
  \country{USA}
  \city{New York City}
}

\author{Rubén González-Sendino}
\email{ruben.gonzalez@eticas.ai}
\orcid{0000-0002-0283-6739}
\affiliation{%
  \institution{Eticas AI}
  \country{Spain}
  \city{Madrid}
}

\author{Paola Vazquez}
\email{paola.vazquez@eticas.ai}
\orcid{0009-0009-6098-2230}
\affiliation{%
  \institution{Eticas AI}
  \country{Ecuador}
  \city{Cuenca}
}

\renewcommand{\shortauthors}{Gemma et al.}

\begin{abstract}

Artificial intelligence (AI) models are increasingly autonomous in decision making, making the pursuit of responsible AI more critical than ever. Responsible AI (RAI) is defined by its commitment to transparency, privacy, safety, inclusiveness, and fairness. But while the principles of RAI are transparent and shared, RAI practices and auditing mechanisms are still incipient. A key challenge is establishing metrics and benchmarks that define performance goals aligned with RAI principles. This paper presents how the ITACA AI auditing platform incorporates demographic benchmarking for AI recommender systems to identify and measure bias. We propose a Demographic Benchmarking Framework to measure populations potentially affected by specific models, set acceptable performance ranges, and guide policymakers and developers. Our approach integrates socio-demographic insights directly into AI systems, reducing bias while also improving overall performance. The main contributions of this study include: 1. Defining control datasets tailored to specific demographics so they can be used in model training to quantify sampling bias; 2. Comparing the overall population with those impacted by the deployed model to identify discrepancies and account for structural bias; and 3. Quantifying drift in different scenarios continuously and as a post-market monitoring of deployment bias.

\end{abstract}

\maketitle

\section{Introduction}

Integrating artificial intelligence (AI) into diverse systems is becoming increasingly easy, and thus pervasive. As AI expands into high-risk areas, such as healthcare, education, banking, and hiring, these systems must adhere to rigorous standards \cite{euaiact}. Beyond existing regulations and laws, the AI community has embraced Responsible Artificial Intelligence (RAI) principles \cite{arrieta2019explainable} to safeguard society from harmful AI impacts.

A core principle within these guidelines is Fairness. Traditionally linked to equality, fairness is defined as “the absence of prejudice or favouritism towards an individual or a group based on its inherent or acquired characteristics” \cite{Baker2021}. However, the evolving social dialogue has shifted the focus toward equity, defined as “understanding and providing to the ones who need the necessary resources to belong to a particular community” \cite{equity_monique}. This broader perspective recognizes that achieving true fairness requires tailored solutions to address historical and structural inequalities.

Achieving fairness in AI is a collaborative effort that involves multiple stakeholders throughout the AI lifecycle: 1) Developers, model creators and system integrators; 2) Policymakers, architects of regulations and laws; 3) Auditors, evaluators responsible for inspecting and assessing AI models to ensure compliance with fairness standards.

The prevalent approaches for detecting bias in AI utilize well-established metrics. These metrics generally concentrate on two key aspects:
\begin{itemize}
    \item Outcome Disparities: Evaluating differences in achieving positive outcomes across groups through measures such as Disparate Impact, Statistical Parity Difference, Equality of Opportunity, Predictive Parity, and Average Odds Difference \cite{conversationai2021}.
    \item Model Performance: Assessing the model’s accuracy by analyzing errors and successes with metrics such as True Positive Rate, False Positive Rate, False Negative Rate Parity, Balanced Error Rate, and Group Calibration \cite{aequitas}.
\end{itemize}

Current fairness evaluation metrics are adept at addressing several emerging biases during model development. As commonly used metrics demonstrate, quantitative measures can often mitigate biases such as cognitive bias, labelling bias, feature selection bias, and evaluation or aggregation bias. However, other biases, such as temporal, sampling, and deployment biases, present significantly greater challenges. Although temporal and deployment biases may sometimes be gauged by detecting drift between production and training data, this evaluation can be misleading if the training data’s distribution is erroneously treated as the ground truth. Moreover, structural issues that stem from deeper societal inequalities frequently remain unchecked.

This paper aims to address bias "moments" outside of the model, measuring bias features and attributes throughout the AI system life-cycle. We do so by adding metrics related to sampling, deployment, and structural bias, which are three moments of bias closely linked to the representation of populations in datasets. \textit{Sampling bias} occurs when the population is not adequately represented in the training set, leading to models that fail to generalize across diverse groups. In contrast, \textit{deployment bias} arises when a model is deployed in an environment with a population distribution that is different from that expected during development.

\textit{Structural bias} occurs when the population lacks the necessary variability in its sociodemographic characteristics, causing membership in a particular group (ethnic, socioeconomic, cultural, educational, professional, etc.) to determine a person’s other attributes or behaviours. This bias arises because some groups may be overrepresented (or underrepresented) in certain fields (e.g., a university major, a profession, or a specific educational environment), so the sample or the actual population does not reflect full diversity. As a result, inferences or conclusions rely excessively on group membership rather than individual characteristics.

Our paper presents a solution to the limited scope of most bias detection metrics used today. Specifically, we introduce a Demographic Benchmarking Framework that evaluates the populations potentially affected by AI models, particularly within recommender systems. This framework enables auditors to measure key performance indicators against acceptable thresholds, supports developers in constructing balanced training datasets and incorporating production performance metrics, and assists policymakers in formulating effective, enforceable regulations.

By providing an external reference for the ideal demographic distribution, our approach encourages stakeholders to look beyond the training data. It ensures that decisions are informed by real-world demographics and dynamics rather than being driven solely by potentially skewed training data. Ultimately, this socio-technical strategy bridges the gap between technical bias detection and broader societal challenges, offering a transformative step toward truly fair and accountable AI.


The paper is structured as follows. Sections \ref{background} and \ref{relatedwork} provide related work that includes the necessary background to understand the approach presented. Section \ref{db} delves into the framework proposed to measure demographic disparities. Section \ref{result} presents the experimental results of evaluating the approach against a real scenario, and Section \ref{discussion} discusses the benefits of this framework for auditors, developers and policymakers. Section \ref{conclusion} concludes and provides an overview of future work. 

\section{Background} \label{background}

Data used to train AI models often contains biases that can lead to discrimination. Although the concept of \textit{bias} is broad, in this paper we define bias as ``the systematic tendency in a model to favour one demographic group/individual over another, which can be mitigated but may well lead to unfairness'' \cite{Fletcher2021,Baker2021}. Biases can be inherited (and later perpetuated) or introduced (and then exacerbated) by AI models. Inherited bias perpetuates existing inequalities in the data structure \cite{zheng2021}, and bias can also be introduced in AI models by assumptions in the implementation of the model that exacerbate discrimination \cite{Vesselinov2019,Akintande2021}.

To address bias, unfairness, and discrimination, several legal frameworks have suggested and mandated the need to take proactive steps to identify and mitigate bias, including bias auditing either as part of broader obligations, such as in the AI Act and the Digital Services Act, or as a stand-alone obligation such as in NYC Law 144. The auditing approach we take follows a procedure that involves: (1) the identification of potential biases that can affect fairness; (2) the selection of metrics to measure how fair AI is being; and (3) the mitigation of the impact produced by these biases \cite{Solans2021,Hajian2016,DeCamp2020}. However, eliminating bias in machine learning and Artificial Intelligence without addressing the pressing concerns about human bias is impossible \cite{Ahmed2021}.

\subsection{Bias}

Three groups could be identified when relating bias to the life-cycle model: data bias, model bias, and deployment bias \cite{pessach2022}. Data bias is related to the dataset selected to train the algorithm, meaning problems of selected samples, features, transformation and labelling. Talking about model bias is linked to exacerbating biases inherent in data or creating new ones that cause learning. Finally, deployment bias is linked to how and where the model is deployed \cite{bias_sr}

\begin{itemize}
    \item  \textit{Data - Cognitive Bias} is a deeply ingrained part of human decision making \cite{Harris2020}, which transfers prejudices to labels \cite{Sharma2020}. Machine learning algorithms use human judgments as training data, so they propagate these biases. 

    \item \textit{Data – Labeled Bias} can arise even when the data is perfectly measured and sampled; for example, by inadvertently reinforcing a stereotype \cite{conversationai2021}.
        
    \item \textit{Data - Feedback loop} is the result of introducing a new discriminatory decision in the data from an algorithm or human decision \cite{Khenissi2020}.

    \item \textit{Data - Behavior Bias} produces distortions from reality or other applications according to user connections, activities, or interactions \cite{Baker2021}. Furthermore, unconscious bias could be produced by \textit{content creation} because the way a child or an adult expresses themselves is different, in the same way as if you compare by sex or race \cite{Badal2021}. Content creation bias is also produced when users are guided by norms or functionalities \cite{Baker2021}, and sometimes these interactions are led by an AI system \cite{DeCamp2020}. Bias production in the future will be affected by unfair systems, generating new data to be used in future learning \cite{Baker2021}.
    
    \item \textit{Data - Sampling Bias} is generally a distortion in \textit{sampled data} that compromises its representatives \cite{Currie2021,mehrabi2021}. In other words, sample bias is produced when the training and testing data do not represent or under-represent a population segment. Therefore, modellers play a critical role in producing data bias by including or discarding data \cite{Akintande2021} and even labelling the data \cite{Baker2021}.

    \item \textit{Data – Feature Bias} can arise when sensitive attributes (race, sex, age, socioeconomic status, education, or neighbourhood) are incorporated into decision-making \cite{Pandey2021}. These attributes are considered illegitimate grounds for decisions like the General Data Protection Regulation (GDPR). However, fairness is not confined to these protected attributes alone: proxy attributes can also infer sensitive information \cite{pessach2022}. In many cases, protected characteristics and target variables are highly correlated, yielding high accuracy but undermining fairness metrics \cite{zheng2021}. Treating correlation as causation can further exacerbate bias \cite{Baker2021}.

    \item \textit{Data – Transformation Bias} emerges when the processes used to transform raw data: normalization, encoding, outlier removal, or feature engineering systematically disadvantage or misrepresent certain subgroups. These procedures can perpetuate or intensify biased results throughout the AI lifecycle by not accounting for how different transformations impact diverse groups \cite{bias_sr}.

    \item \textit{Model - Aggregation Bias} are produced in model training. Typically, models learn a correct statistical pattern in favour of the majority over minorities \cite{zheng2021,mehrabi2021}, leading to aggregation bias that amplifies the disparities between different examples in data samples \cite{conversationai2021}. At this point, the model that learns from generated experience, such as reinforcement learning, could become biased over time \cite{DeCamp2020}.

    \item \textit{Model - Evaluation Bias} occurs when the selected metrics are not appropriate, for example, using general vs. subgroup accuracy \cite{Hou2021}, and when the representation in the test data does not reflect reality. 

    \item \textit{Deployment bias} may occur in the deployment and use of the model. The algorithm makes decisions based on patterns learned from the data. Therefore, \textit{the deployment} of a model in a different scenario concerning data could lead to unfair results \cite{Fletcher2021}. 

\end{itemize}

The forms of bias discussed in the literature do not fully account for societal structural bias. Most commonly recognized biases focus on data or model shortcomings. By introducing the concept of Structural Bias, this paper highlights how societal structures intersect with technical dynamics, and proposes a clear path towards the development of socio-technical evaluation metrics for AI systems.

\subsection{Metrics}

The metrics commonly used to measure fairness are associated with the comparison of privileged (PG) and underprivileged (UG) groups. There are also metrics to compare individuals, but they are less popular \cite{Zhang2020}. The metrics have a valid range that has previously been defined in the literature \cite{Bellamy2018AIF3}.  

These fairness metrics operate by juxtaposing two distinct groups. Converting sensitive features into binary variables is a prevalent approach, aligning various categories within these features with privileged or underprivileged groups \cite{Bellamy2018AIF3}. In addition, an automated algorithm is available to identify the privileged group, enabling a comprehensive comparison among all groups and selecting the optimal group as a reference point \cite{aequitas}. The most common metrics used to measure fairness are:

\begin{itemize}
    \item Equal Opportunity Difference \cite{Stevens2020}. Measures the difference in true positive rates (TPR) between the underprivileged and privileged groups. The ideal value is 0; in this study, the interval between -0.1 and 0.1 will be fair.
          \begin{equation} \label{tpr}
        	\textrm{TPR}  = \big[{\frac{\textrm{True Positive (TP)} }{\textrm{TP} + \textrm{False Negative (FN)}}}\big]
        	\end{equation}
          \begin{equation} \label{eod}
        	\textrm{EOD}  = \textrm{TPR}_{\textrm{UG}} - \textrm{TPR}_{\textrm{PG}}  
        	\end{equation}
    \item Odds Difference \cite{Ahmed2021}. Calculates the difference between the underprivileged and privileged groups in false positive rates (FPR) and true positive rates (TPR). The ideal value is 0; in this study, the interval between -0.1 and 0.1 will be fair.
        \begin{equation} \label{tpr2}
        	\textrm{FPR}  = \big[{\frac{\textrm{False Positive (FP)} }{\textrm{FP} + \textrm{True Negative (TN)}}}\big]
        	\end{equation}
          \begin{equation} \label{eod2}
        	\textrm{OD}  = \big(\textrm{FPR}_{\textrm{UG}} - \textrm{FPR}_{\textrm{PG}}\big)   + \big(\textrm{TPR}_{\textrm{UG}} - \textrm{TPR}_{\textrm{PG}}\big)  
        	\end{equation}
    \item Statistical Parity Difference \cite{Sharma2020}. It calculates the difference in the probability of favourable results (Predicted as Positive (PPP)) between the underprivileged and privileged groups. The ideal value is 0; in this study, the interval between -0.1 and 0.1 will be fair.
          \begin{equation} \label{ppp}
        	\textrm{PPP}  = \big[{\frac{\textrm{TP} + \textrm{FP}}{\textrm{Total Population (N)}}}\big]
        	\end{equation}
          \begin{equation} \label{spd}
        	\textrm{SPD}  = \textrm{PPP}_{\textrm{UG}} - \textrm{PPP}_{\textrm{PG}}  
        	\end{equation}
    \item Disparate Impact \cite{Alam2020}. Compares the proportion of individuals who receive a positive output for The underprivileged and the privileged. The ideal value is 1. In this study, the interval between 0.8 and 1.2 will be fair.
             \begin{equation} \label{di}
            	\textrm{DI}  = \frac{\textrm{PPP}_{\textrm{UG}}}{\textrm{PPP}_{\textrm{PG}}}  
            	\end{equation}

As commonly used metrics show, many previously mentioned biases (cognitive bias, labelling bias, feature selection bias, and evaluation or aggregation bias) can often be effectively addressed. However, biases like temporal, sampling, or deployment bias remain more challenging to measure and evaluate. Although temporal or deployment bias can sometimes be gauged by detecting drift between production and training data, this assessment may be skewed if the training data’s distribution is treated as the ground truth.

Moreover, structural issues may arise and remain overlooked. Therefore, this paper aims to examine how population and deployment biases are evaluated and to introduce a new concept—structural bias—that can help identify and mitigate deeper societal challenges.

\end{itemize}
\section{Related Work} \label{relatedwork}

In recent years, the study of bias in artificial intelligence systems has gained substantial attention, with researchers focusing on fairness, accountability, and transparency. This section reviews key contributions to detecting bias in AI systems, demographic representation, and fairness metrics.

\subsection{Fairness Metrics and Bias Detection}

Barocas et al. \cite{Barocas2016BigDD} provided one of the first comprehensive analyses of how discrimination manifests in machine learning systems. This demonstrates how bias can be accidentally encoded through data collection, feature selection, and modelling choices. Their work established a critical framework for understanding algorithmic discrimination that continues to influence the field.

Expanding on these foundations, Chouldechova and Roth \cite{Chouldechova/3376898} presented a detailed analysis of the mathematical relationships between different fairness criteria, demonstrating that certain fairness metrics are fundamentally incompatible. This mathematical framework helped explain why achieving multiple fairness objectives simultaneously often presents inherent trade-offs.

\textbf{Demographic representation in data sets}

A landmark study by Buolamwini and Gebru \cite{buolamwini18a} exposed significant accuracy disparities between demographic groups in commercial facial analysis systems. Their research introduced the Gender Shades methodology for evaluating bias in AI systems and highlighted how underrepresentation in training data leads to discriminatory outcomes.

Holstein et al. \cite{Holstein_2019} conducted extensive interviews with industry practitioners developing fair ML systems, revealing critical gaps between academic fairness research and real-world needs. Their findings emphasized the necessity for practical tools to assess demographic representation throughout the model development lifecycle.

Recent advances in demographic fairness research have brought new perspectives to the field. Wang et al. \cite{Wang2021} introduced an innovative framework for dynamic fairness assessment in early 2024, addressing how demographic shifts over time impact AI system performance. Their work demonstrates the importance of continuous monitoring and adaptation of fairness metrics as population distributions change.

\textbf{Regulatory frameworks and Auditing}

On the other hand, Raji and colleagues \cite{Raji/3351095.3372873} introduced SMACTR, a structured framework for internal algorithmic auditing that has significantly influenced how organizations evaluate AI systems for potential biases. This work provided concrete guidance for implementing algorithmic audits in practice.

In automated hiring, Raghavan et al. \cite{Raghavan/3351095.3372828} analyzed various employment algorithms and their impact on workforce diversity. Their research influenced recent regulatory initiatives, including New York City's Local Law 144 governing automated employment decision tools.

\textbf{Benchmarking approaches}

Recent work by Mehrabi et al. \cite{Mehrabi/3457607} proposed a comprehensive taxonomy of fairness metrics and evaluation methodologies. Their framework emphasizes the importance of context-specific benchmarking and introduces novel metrics for assessing intersectional fairness.

While many contributions have examined the issue of bias and explored its connection to structural and social dynamics, technical approaches to bias detection often rely on narrow definitions of bias. Our approach and platform embrace socio-technical complexity to generate data on bias that accurately reflects a model's broader impact on societal fairness.

\section{Demographic Benchmarking} \label{db}

This section introduces demographic benchmarking, which defines the expected proportion of representation for each group in a production environment. For example, for the protected attribute sex, groups such as males and females are typically expected to be represented equally, with a fifty-fifty distribution. In contrast, the representation of different races or the intersection of sex and race is often less commonly known. For this reason, it is crucial to establish this reference to understand and evaluate how these groups should be defined.

Creating demographic benchmarking is not a trivial task. It requires continuous improvement and involves two key steps: training and production.

\begin{itemize} 
    \item Training: During this step, demographic benchmarking is initially created. This involves considering the population as all individuals potentially impacted by the model. For instance, in a hiring system, the analysis might focus on the distribution of the active workforce population. 
    
    \item Production: This step collects feedback about the actual population engaged with the model. This stage is particularly significant because it serves as a critical inflexion point to determine whether the real-world distribution aligns with the expectations set by the demographic benchmarking or if the impacted population differs from expected.
\end{itemize}

\begin{figure*}[ht!]
  \centering
  \includegraphics[width=\linewidth]{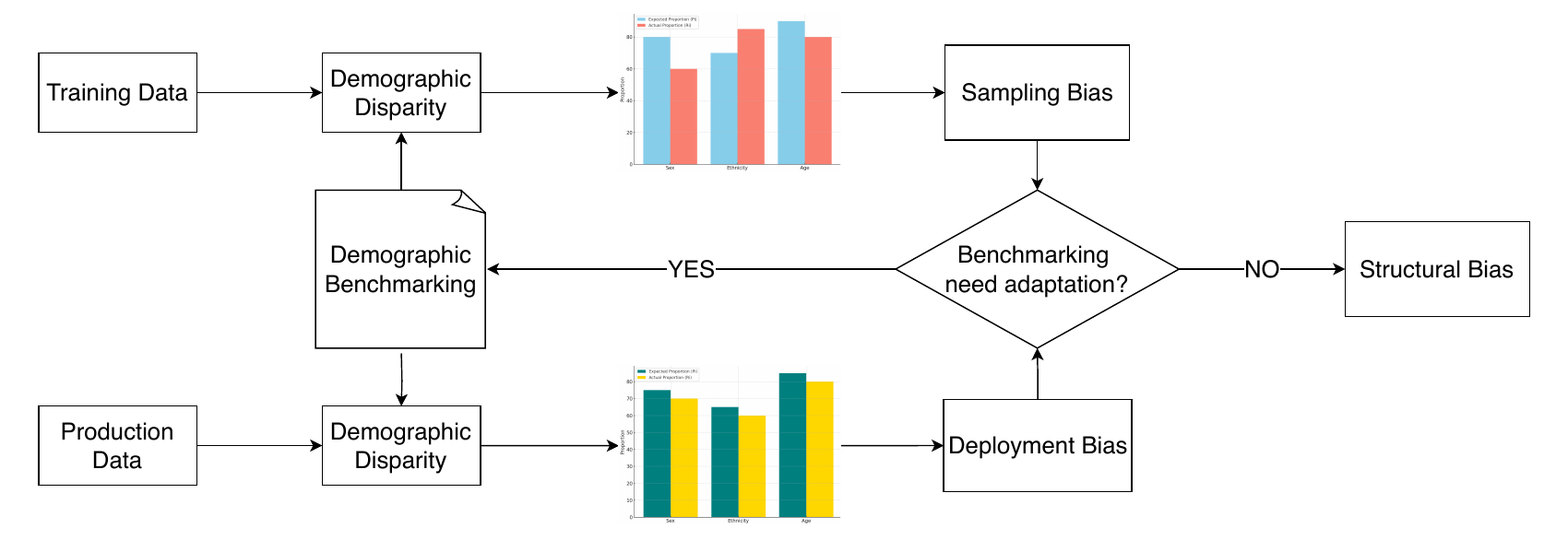}
  \caption{Flowchart illustrating the role of metrics in ensuring equal proportion against reality and usability across development and production.}
  \label{fig:flowchart}
\end{figure*}

Figure \ref{fig:flowchart} illustrates the applicability of the metric across both Training and Production data. This enables the use of these metrics in the production environment, effectively overcoming the limitations of certain metrics that require true labels to monitor bias. The monitoring sampling, deployment, and structural bias results provide valuable insights for auditors, developers, and policymakers, enabling them to effectively track and address bias.

Demographic disparity (DD) is quantified to calculate and measure demographic issues as shown in Eq. \ref{bench_equation}. It is defined as the difference between the expected proportion (\( P_i \)) and the actual proportion in the dataset (\( R_i \)), where \textit{i} represents a specific demographic group (e.g., male, female, or a combination of attributes such as race and sex). This measures the discrepancy between each group's anticipated and observed representation within the dataset.

To extend the analysis across all demographic groups, the Total Demographic Disparity (TDD) is calculated as shown in Eq. \ref{total_disparity}. This metric aggregates the absolute discrepancies for all demographic groups, providing a comprehensive measure of the overall disparity in the dataset. 

While TDD captures the extent of representation imbalances, it does not account for variations in the relative size of each group. To address this, the Normalized Demographic Disparity (NDD), defined in Eq. \ref{normalized_disparity}, introduces a proportional adjustment by dividing each group's discrepancy by its expected proportion and averaging across all groups. NDD provides a more slight view, highlighting relative imbalances and allowing for comparisons across datasets with differing demographic distributions.

\begin{equation} \label{bench_equation}
\textrm{Demographic Disparity} = \textrm{Expected Proportion} (P_i) - \textrm{Actual Proportion} (R_i)
\end{equation}

\begin{equation} \label{total_disparity}
\textrm{Total Demographic Disparity} = \sum_{i=1}^n \left| P_i - R_i \right|
\end{equation}

\begin{equation} \label{normalized_disparity}
\textrm{Normalized Demographic Disparity} = \frac{1}{n} \sum_{i=1}^n \frac{\left| P_i - R_i \right|}{P_i}
\end{equation}

In addition, the following metrics have been added to evaluate the distribution when the decision (D) represents positive results. This decision represents the true label in training and the model's output in production. The added value of this metric is the possibility of measuring whether the model's results follow equitable output concerning the expected distribution.

\begin{equation} \label{bench_equation_d1}
\textrm{DD Positive Decision (DDP)} = P_i(D = 1) - R_i
\end{equation}

\begin{equation} \label{total_disparity_d1}
\textrm{TDD Positive Decision (TDDP)} = \sum_{i=1}^n \left| P_i(D = 1) - R_i \right|
\end{equation}

\begin{equation} \label{normalized_disparity_d1}
\textrm{NDD Positive Decision (NDDP)} = \frac{1}{n} \sum_{i=1}^n \frac{\left| P_i(D = 1) - R_i \right|}{P_i(D = 1)}
\end{equation}

\section{Use Case: Hiring New York City} \label{result}

\begin{table}[h!]
\resizebox{1\textwidth}{!}{%
\centering
\begin{tabular}{|l|l|l|r|r|r|}
\hline
\textbf{Code} & \textbf{Race} & \textbf{Sex} & \textbf{Population (NYC)} & \textbf{Percentage (NYC)} & \textbf{Percentage (USA)}\\ \hline
1 & American Indian and Alaska Native alone & Female & 7942 & 0.11 & 0.31 \\ \hline
2 & American Indian and Alaska Native alone & Male & 7292 & 0.10 & 0.30 \\ \hline
3 & Asian alone & Female & 613474 & 8.45 & 3.77\\ \hline
4 & Asian alone & Male & 542064 & 7.47 & 3.32 \\ \hline
5 & Black or African American alone & Female & 816201 & 11.24 & 6.97 \\ \hline
6 & Black or African American alone & Male & 657805 & 8.92 & 5.97 \\ \hline
7 & Hispanic or Latino & Female & 1051246 & 14.48 & 10.45 \\ \hline
8 & Hispanic or Latino & Male & 924348 & 12.73 & 10.03 \\ \hline
9 & Native Hawaiian and Other Pacific Islander alone & Female & 1596 & 0.02 & 0.11 \\ \hline
10 & Native Hawaiian and Other Pacific Islander alone & Male & 1182 & 0.02 & 0.11\\ \hline
11 & Some Other Race alone & Female & 47297 & 0.65 & 0.25 \\ \hline
12 & Some Other Race alone & Male & 43784 & 0.60 & 0.25 \\ \hline
13 & Two or More Races & Female & 127611 & 1.76 & 1.85 \\ \hline
14 & Two or More Races & Male & 103630 & 1.43 & 1.64 \\ \hline
15 & White alone & Female & 1198658 & 16.51 & 28.11\\ \hline
16 & White alone & Male & 1126478 & 15.51 & 26.56\\ \hline
\end{tabular}%
}
\caption{Demographic Data by Race, Sex, Value, and Percentage for New York City (NYC) and United States (US). All race categories presented exclude individuals identifying as Hispanic or Latino (nor Hispanic or Latino).}
\label{tab:demographic_data}
\end{table}

\begin{table}[h!]
\resizebox{1\textwidth}{!}{%
\centering
\begin{tabular}{|l|l|l|l|r|r|r|r|}
\hline
\textbf{Code} &  \textbf{Percentage (USA)}&\textbf{\% ADP} & \textbf{Positive \% ADP} & \textbf{\% RippleMatch} & \textbf{Positive \% RippleMatch} & \textbf{\% SheppardMullin} & \textbf{Positive\% SheppardMullin}\\ \hline
1 &  0.31 &0.2 & 0.2 & 0.21 & 0.16 & 0.07 & 0.06 \\ \hline
2 &  0.30 &0.13 & 0.13 & 0.44 & 0.33 & 0.07 & 0.09 \\ \hline
3 &  3.77&5.6 & 5.34 & 24.82 & 24.7 & 11.23 & 10.70 \\ \hline
4 &  3.32 &7.16 & 6.34 & 36.16 & 34.84 & 6.75 & 6.99 \\ \hline
5 &  6.97 &15.84 & 15.64 & 3.22 & 3.31 & 5.70 & 5.90 \\ \hline
6 &  5.97 &9.10 & 8.80 & 6.19 & 6.38 & 2.89 & 3.06 \\ \hline
7 &  10.45 &9.08 & 9.09 & 1.82 & 1.81 & 6.31 & 7.06 \\ \hline
8 &  10.03 &8.30 & 8.23 & 4.73 & 5.05 & 5.20 & 5.38 \\ \hline
9 &  0.11 &0.12 & 0.11 & 0.03 & 0.04 & 0.02 & 0.03 \\ \hline
10 &  0.11&0.10 & 0.10 & 0.09 & 0.08 & 0.04 & 0.05 \\ \hline
11 &  0.25 &0.0 & 0.0 & 0.0 & 0.0 & 0.0 & 0.0 \\ \hline
12 &  0.25 &0.0 & 0.0 & 0.0 & 0.0 & 0.0 & 0.0 \\ \hline
13 &  1.85 &2.24 & 2.28 & 5.19 & 5.83 & 2.79 & 2.78 \\ \hline
14 &  1.64 &1.44 & 1.43 & 2.31 & 2.49 & 2.05 & 2.00 \\ \hline
15 &  28.11&20.99 & 22.13 & 5.23 & 5.51 & 28.18 & 29.77 \\ \hline
16 &  26.56&19.71 & 20.18 & 9.55 & 9.48 & 28.69 & 26.14 \\ \hline
\end{tabular}%
}
\caption{Demographic Data by Race, Sex, Value, and Percentage for total and positive samples. All race categories presented match using code with categories presented on Table \ref{tab:demographic_data}}
\label{tab:demographic_data_case}
\end{table}

This section explains how demographic benchmarking is created and used. To illustrate its practical application, we focus on a hiring scenario in New York City, selected due to New York City Local Law 144 (LL144), which regulates automated employment decision tools (AEDTs) to ensure fair candidate screening and promotion processes.

Table \ref{tab:demographic_data} shows the demographic benchmarking approach, which provides population distributions across race and sex. These data were obtained from the U.S. Census Bureau’s 2020 census. To align with a hiring context, the dataset was filtered to include only the active workforce—specifically, individuals aged 16 and older.

A real AEDT was used as a reference point to evaluate how demographic benchmarking supports system auditing and analysis. By examining public audit reports published under LL144, we gathered the number of samples for each sex and race category, allowing us to derive the population distribution within these systems. These audit reports were obtained from the following repository: \href{https://github.com/aclu-national/tracking-ll144-bias-audits?tab=readme-ov-file}{ACLU National: Tracking Local Law 144 Bias Audits}.

Three algorithms, RippleMatch, Sheppard Mullin, and ADP, have been selected from this repository. Unfortunately, the exact data used to train these models is not disclosed. Although LL144 is for New York City candidates, the opaque audits regarding candidates' location data required us to compare results against the broader U.S. population distribution.

Table \ref{tab:demographic_data_case} shows the distribution of total candidates and those who received a positive outcome. The “Some other race” category is omitted because it is not required by law. The results indicate that each tool affects distinct populations. For example, ADP exhibits a notable deviation in the Black category compared to the benchmark, while RippleMatch shows an elevated representation of Asian applicants and a lower representation of White applicants.

These discrepancies may be due to several factors, such as the types of positions posted on each platform, the companies using the tools, and the visibility of the platforms to potential applicants. Thus, the expected population differs from the impact population; this requires revision to improve the evaluation process. Moreover, LL144 does not mandate that the audit data come exclusively from New York City candidates, making it difficult to establish a perfectly localized benchmark. 

Despite these limitations, demographic benchmarking proves valuable in assessing a model’s behaviour and performance without requiring access to the original training set. Examining how these algorithms operate in production can help us gain insights into their socio-technical implications.

In addition, these metrics are useful throughout the AI lifecycle—from model training to post-deployment monitoring—by providing developers and policymakers with a reference population that enables more accurate and equitable evaluations across various AI systems.

\section{Discussion} \label{discussion}

This section delves into three key types of bias underscored by our demographic benchmarking approach: sampling bias, deployment bias, and structural bias. Each of these biases can arise at different stages of the AI lifecycle, ranging from how datasets are created to how models learn from those datasets to the real-world contexts in which models are used. The benefits of this framework are discussed from three distinct viewpoints: developers, policymakers, and auditors.

\subsection{Developers}

Demographic benchmarking is a methodology for evaluating machine learning models throughout the training and production phases. It references expected demographic distributions and guides decision-making at every stage of a model’s life cycle.

Demographic benchmarking helps developers construct datasets that accurately represent the target population during training. Addressing sampling bias ensures the training data aligns with the diversity of the model's intended target groups.

In the production phase, demographic benchmarking allows developers to monitor the real-world populations interacting with the model. This monitoring helps identify deployment bias whenever the population differs from the expected distribution established during training.

By applying demographic benchmarking in both phases, developers can tackle critical biases and maintain fairness in their models. This practice ensures that models perform reliably, adapt to evolving conditions, and adhere to society and operational standards.

\subsection{Policymakers}

Demographic benchmarking is exactly key for policymakers. It provides a systematic framework for auditing algorithms and evaluating fairness. This framework enables policymakers to spot discrepancies and highlight potential risks, laying the groundwork for more robust regulations and auditing practices.

Addressing gaps in current regulations—like those in the EU AI Act—remains a pressing challenge. Although the Act emphasizes the need for accurate representation in datasets, it does not provide explicit guidelines for measuring or implementing those requirements. Demographic benchmarking fills this void by introducing quantifiable indicators of representation and fairness, offering a practical pathway toward meeting regulatory objectives.

New York City’s Local Law 144 presents additional challenges, especially because it excludes demographic groups representing less than 2\% of the dataset from impact-ratio calculations. This exclusion is problematic for groups like Native Hawaiian or Pacific Islander and Native American or Alaska Native populations, whose real-world representation in New York City is already below 2\%.

By integrating demographic benchmarking into regulatory frameworks, policymakers can address these shortcomings and set more comprehensive fairness and accountability standards. Ultimately, this approach ensures that all populations, irrespective of size, are represented and protected in AI-driven systems, fostering equitable and inclusive outcomes.

\subsection{Auditors} 

Auditors play an essential role in evaluating machine learning systems. They are impartial entities tasked with ensuring fairness and accountability. Their unique position allows them to bridge gaps between development and production environments and across different organizations, thus guiding the effective definition and application of regulations.

By leveraging a robust demographic benchmarking framework, auditors gain an additional reference point for validating datasets and outcomes from a societal perspective. This external benchmark helps them assess whether AI systems adequately reflect real-world demographics and address potential inequities that might not be apparent when focusing solely on technical metrics.

Furthermore, auditors facilitate developers' adoption of demographic benchmarking. They offer guidance on crafting representative datasets and monitoring real-world performance, sharing best practices and lessons learned across diverse audits. This approach ensures that machine learning models are built on solid foundations, prioritizing fairness from the outset.

In conclusion, auditors guarantee that AI systems are developed and deployed ethically and transparently. By grounding their evaluations in demographic benchmarking, they reinforce public trust in AI, reduce concerns about fairness and effectiveness, and ultimately pave the way for wider societal acceptance of these technologies.

\section{Conclusions and Future Work} \label{conclusion}

Measuring all the factors that influence AI systems is a complex task. Currently, most evaluations focus on model validation using true labels and success rates. However, these methods are largely limited to examining training data and model outputs, lacking external information that could enrich the assessment. The demographic benchmarking framework introduced in this paper addresses this gap by incorporating societal data.

Through testing, demographic benchmarking has proven to be a valuable resource for providing external information about the population’s distribution. By indicating the expected population likely to be impacted by the model, it enables the evaluation of three major biases that have traditionally been difficult to measure:

\begin{itemize}
    \item Sampling Bias. Constructing an accurate training dataset is challenging when there is no clear understanding of the target population. Analyzing the expected population for each model offers developers a reference point during the design phase, guiding the creation of training data that more faithfully reflects societal diversity.

    \item Deployment Bias. While drift detection techniques have been used in the literature to address this bias, comparisons have largely been limited to production data versus training data—thereby inheriting any biases in the training dataset. By incorporating demographic benchmarking, production data can be assessed against the anticipated real-world population, providing insights into the operating environment and allowing models to better adapt to societal needs.

    \item Structural Bias. This paper also introduces a new bias type, Structural Bias, which captures the extent to which society may be inherently unfair. Existing structural barriers can significantly affect future outcomes for certain groups. By quantifying these societal imbalances, demographic benchmarking helps policymakers identify underprivileged groups and guides companies in determining where equity-focused interventions are most urgently needed.
\end{itemize}

In our future work, we will continue to work to refine a socio-technical approach to AI auditing and bias measurement, testing and implementing comprehensive benchmarking data across diverse fields and use cases, and making benchmarking data available. These benchmarks enable us to drive innovation in AI auditing, help developers integrate socio-technical metrics in production, and empower policymakers with evidence-based insights to shape regulation.  

As pioneers in the AI auditing industry at Eticas AI, we believe that auditors play a crucial role in aggregating insights from varied applications to create robust, reusable demographic benchmarks. If AI auditors incorporate robust, socio-technical benchmarks, they can create incentives for the industry to incorporate better bias detection and mitigation measures and metrics in production, thus releasing AI systems that do a better job at controlling for bias in credible ways.

\bibliographystyle{ACM-Reference-Format}
\bibliography{sample-base}

\end{document}